\documentclass[11pt, oneside]{article}   	
\usepackage[a4paper, total={6in, 8in}]{geometry}               		
\usepackage{graphicx}				
\usepackage{amssymb}
\usepackage{gensymb}
\usepackage[toc,page]{appendix}
\usepackage{graphicx}
\usepackage{subcaption}
\usepackage{placeins}
\usepackage{float}

\title{Experimentally Validated Simulations of 50~$\mu$m X-ray PIV Tracer Particles}
\author{Jason T. Parker \& Simo A. M\"akiharju}

\begin{document}
\maketitle

\begin{abstract} \label{sec:abstract}
We evaluate Beer-Lambert (BL) ray-tracing and Monte Carlo N-Particle (MCNP) photon tracking simulations for prediction and comparison of X-ray imaging system performance. These simulation tools can aid the methodical design of laboratory-scale X-ray particle image velocimetry (XPIV) experiments and tracer particles by predicting image quality. Particle image signal-to-noise ratio (SNR) is used as the metric of system performance. Simulated and experiment data of hollow, silver-coated, glass sphere tracer particles (AGSF-33) are compared. As predicted by the simulations, the AGSF-33 particles are visible with a SNR greater than unity in 100~ms exposure time images, demonstrating their potential as X-ray PIV or particle tracking velocimetry (XPTV) tracers. The BL approach predicts the image contrast, is computationally inexpensive, and enables the exploration of a vast parameter space for system design. MCNP simulations, on the other hand, predict experiment images slightly more accurately, but are more than an order of magnitude more computationally expensive than BL simulations. For most practical XPIV system design applications, the higher computational expense of MCNP is likely not justified by the modest accuracy improvement compared to BL.
\end{abstract}

\section{Introduction}
Particle Image Velocimetry (PIV) \cite{m_raffel_particle_2018} and particle tracking velocimetry (PTV) can both provide space- and time-resolved velocity field measurements that are readily compared to theory and simulation. With no probe to disturb the flow field, these are powerful experimental techniques that are widely used in the fluid dynamics community. In what follows we focus on PIV, although much of this work would also apply to PTV.

In typical 2D PIV, one illuminates tracer particles, usually with a laser light sheet. Images of the tracer particles in the fluid are captured by a digital camera. The images are divided into windows, which are then cross-correlated to corresponding windows in a sequential frame to determine the displacement of the tracer particles in each window. Since the frame rate is known, the average velocity of the tracer particles within a window can be calculated. Doing this for each window in the frame reveals the velocity field, resolved to the size of the windows. If the tracer particles are sufficiently small and neutrally buoyant, they are assumed to follow the flow field accurately without affecting the flow field itself. In order for PIV to work, optical access is required for the laser light to reach the tracer particles, and the scattered light must be able to reach the camera sensor. However, for many flows, optical access is impossible due to the opacity of the fluid itself or the surrounding material. For example, in gas-liquid flows visible light is refracted at numerous curved phase boundaries, making the system opaque in the visible spectrum.

X-rays, on the other hand, have a refraction index near unity, and can pass through materials that are opaque at visible wavelengths. As a result, XPIV would not have the optical access limitations of standard PIV. The impact of XPIV cannot be understated. The whole arsenal of three decades of PIV algorithm development could be applied to previously inaccessible systems such as biological flows \cite{antoine_flow_2013, kim_x-ray_2006, jamison_x-ray_2012, krebs_initial_2020}, multiphase flows \cite{ganesh_bubbly_2016, makiharju_time-resolved_2013, yoon_image_2018}, and opaque internal flow channels \cite{park_x-ray_2016, lappan_x-ray_2020}.

In 2003, Lee and Kim demonstrated quantitative XPIV for the first time at a synchrotron by measuring the depth-integrated flow profile in a cylindrical pipe \cite{lee_x-ray_2003}. Since then, research to extend XPIV capabilities \cite{dubsky_computed_2009, fouras_three-dimensional_2007} has been mostly limited to synchrotrons due to the formidable challenges in obtaining a sufficient signal-to-noise ratio (SNR) given the reduced attenuation contrast of micron-sized particles and comparatively dim in-lab X-ray sources. Limiting XPIV to synchrotrons hinders the pace of research by imposing time, cost, and location limitations. Furthermore, synchrotrons typically have illuminated areas on the order of millimeters and place constraints on experiment geometry, materials, and controls. It is clear that developing laboratory-scale XPIV is imperative for the technique to proliferate. 

In order to achieve particle image contrast, previous in-lab XPIV experiments have relied on particles that are either density mismatched with the surrounding fluid, are on the order of a millimeter in size, or both \cite{lappan_x-ray_2020, lee_development_2009, heindel_x-ray_2008}. Particles that are not neutrally buoyant bias the velocity measurements, and large particles – even if nominally density matched – limit the spatial resolution and may not trace the flow if the Stokes number is greater than unity. Hence for all but narrow range of experimental conditions, data from such measurements can only be taken as qualitative. To the best knowledge of the authors, there is no prior laboratory-scale XPIV study with neutrally buoyant (in water) tracer particles on the order of tens of micrometers.

Poelma \cite{poelma_measurement_2020} discusses additional challenges to laboratory XPIV, such as shutter speeds and beam width. Some proposed laboratory-scale XPIV systems offer solutions that are not scalable to higher frame rates, which limits their applicability to flow systems of interest. For example, Lee et al. \cite{lee_development_2009} developed a system with a rotating lead mechanical shutter and achieved a frame rate of 4.5~Hz and a 9~$\mu$m pixel pitch. However, in such a system, the frame rate, field of view, and exposure time are all coupled, making it difficult to scale.

Although the photon flux of laboratory X-ray sources is increasing \cite{stock_liquid-metal-jet_2014}, tracer particles can be detected with more sensitive, ``noiseless" imagers. Photon-counting detectors (PCDs) have seen rapid development in recent years, opening the door to X-ray images with lower noise levels. Unlike traditional X-ray detectors such as charge coupled device detectors, PCDs count individual photons. While traditional X-ray detectors have Gaussian thermal dark noise, PCDs have no dark noise besides cosmic radiation. By lowering the noise level, PCDs can capture higher SNR X-ray images.

Thanks to modern manufacturing processes, it is now feasible to develop $O$(10 $\mu m$) custom tracer particles. However, this is a costly proposition. For a given experimental geometry, X-ray source, and detector, the properties of an ideal tracer may be different. Image simulation tools are needed to methodically and economically improve these expensive XPIV systems. Polychromatic, attenuation-based imaging is highly nonlinear; it is difficult to intuit what effects a design change will have. For example, it may be unclear if a tracer particle made from new materials would improve image contrast when a source with specific characteristics peaks is used. Numerically exploring the parameter space for a variety of source, domain, detector, and tracer properties can be advantageous when developing laboratory-scale XPIV systems.

In this paper, we explore two X-ray simulation methods: Monte Carlo N-Particle photon transport (MCNP) and Beer-Lambert ray tracing (BL). We compare the simulated images to images recorded with a commercially available PCD, polychromatic X-ray source, and tracer particle. Lastly, we utilize the now experimentally validated simulation tools to predict the SNR of X-ray images as a proof-of-concept for the design utility of the simulation tools. The experiments and simulations demonstrate the feasibility of localizing 50~$\mu$m tracers for XPIV using laboratory-scale equipment.

\section{Experimental Setup} \label{sec:expSetup}
\subsection{X-ray Imaging Setup}\label{ssec:imaging}
\begin{figure}
    \centering
    \includegraphics[width=0.5\textwidth]{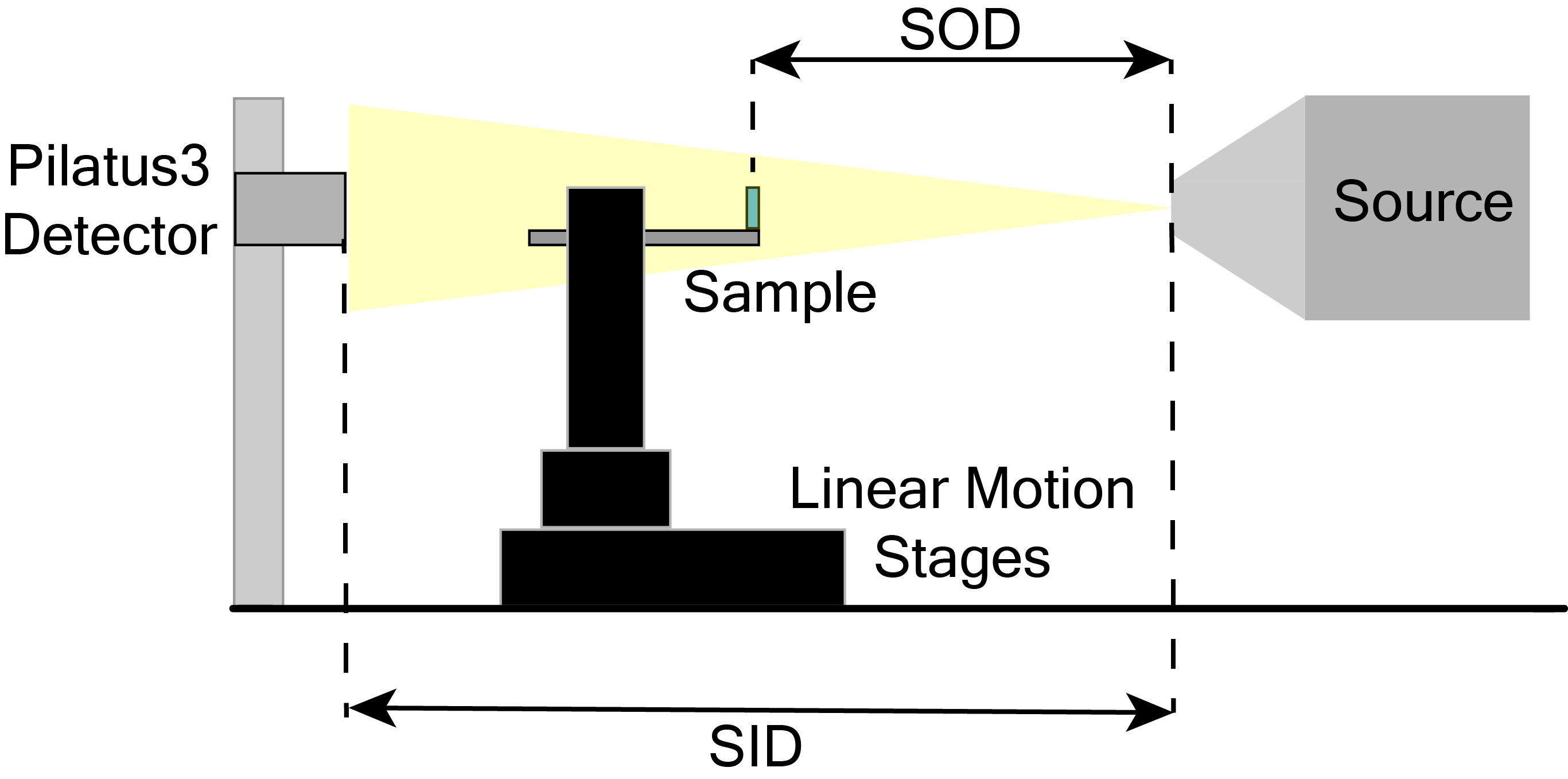}
    \caption{A diagram of the X-ray imaging system. The SOD and SID are 1~cm and 47.6~cm, respectively.}
    \label{fig:expSetup}
\end{figure}

A schematic of the experimental setup is shown in figure \ref{fig:expSetup}. The X-ray source used for these experiments is an YXLON FXE225.99 TwinHead. TwinHead refers to the ability to swap the source heads between a transmission and directional head. This source is capable of achieving up to 225~kV with either head, with a maximum tube power of 64~W with the transmission head and 320~W for the directional head. The directional head is brighter, but sacrifices sharpness due to a larger focal spot. Blurring from the directional head increases the size of the minimum detectable detail from $<$500~nm for the transmission head to $<$3~$\mu$m \cite{noauthor_yxlon_nodate}. The directional source head has a water-cooled tungsten target and a millimeter-thick aluminum window. For our experiments, the directional head was used for higher flux.

The imager used is a Dectris Pilatus3 X 100K--M cadmium telluride (CdTe) detector with a 1~mm thick CdTe panel, 487x195 (172~$\mu$m)$^2$ pixels, and a single photon energy threshold. For the present study the energy threshold is set to 20~keV. The detector is capable of counting approximately $10^6$ photons per pixel per second without correction, which introduces a limit for image quality. The detector can record images at a rate of up to 500 frames per second, but for these experiments it is operated at 10~Hz with an exposure time of 100~ms. The detector temperature was maintained with a chiller at 23.0~$\pm 0.1\degree$C and a continuous dry nitrogen gas purge was used to prevent the buildup of humidity within the detector. The raw data is recorded by the Pilatus3 detector. Prior to recording the images of the particles, a series of flat field images are recorded to generate a correction map. This correction map eliminates detector artifacts and dead pixels when applied to a data image. 

Three ThorLabs NRT150 linear stages control the placement of the object to be imaged. Each stage has a 150~mm range of motion, 0.1~$\mu$m minimum incremental motion, and a calibrated on-axis accuracy of 2~$\mu$m. Samples are held in place with 3D printed mounts that are bolted to the linear stages. When imaging the particles on a glass slide, as shown in figure \ref{fig:expSetup}, the source-to-object distance (SOD) is 1~cm to maximize the geometric magnification effect. The source-to-imager distance (SID) is 47.6~cm, so the geometric magnification is 47.6 times.

\subsection{Tracer Particle}\label{ssec:particleChoice}
For quantitative XPIV, the tracer particle must be neutrally buoyant and have a Stokes number much less than unity \cite{m_raffel_particle_2018}. The AGSF-33 particles from Potters Industries LLC are selected as a potential candidate for XPIV because they are nominally neutrally buoyant in water and their silver coating enhances X-ray attenuation contrast. With a nominal diameter of 50~$\mu$m, AGSF-33 particles will have a Stokes number less than unity for a wide range of flows of interest and can provide reasonable spatial resolution in XPIV.

The particle geometry should be known to accurately simulate a particle. While the actual shape of a particle may deviate from a perfect sphere, we will assume spherical symmetry so multi-layer particles of arbitrary composition can be easily modeled. In the case of AGSF-33, the particles are known to be hollow Pyrex spheres that have been coated with silver that accounts for approximately 33\% ($\gamma = 0.33$) of the particle mass. The glass wall and silver coating thickness can be calculated by assuming neutral buoyancy in water and using the manufacturer's nominal outer diameter, 50~$\mu$m. Let $R_0$ be the inner radius and $R_1$ be the outer radius of the Pyrex wall, and $R_2$ the outer radius of the silver coating. In this case, $R_2=D/2=25~\mu$m. Then, given the densities of air, Pyrex, silver, and known value of $\gamma$, $R_1$ and $R_0$ are readily calculated by equations \ref{eq:r1} and \ref{eq:r0}, respectively.

\begin{equation}\label{eq:r1}
    R_1^3 = \frac{\rho_{silver} - \gamma \rho_{average} }{\rho_{silver}} R_2^3
\end{equation}
\begin{equation}\label{eq:r0}
    R_0^3 = \frac{ (1-\gamma)\rho_{average} R_2^3 - \rho_{pyrex} R_1^3 }{ \rho_{air} - \rho_{pyrex} },
\end{equation}
where $\rho_i$ is the density of material $i$. For AGSF-33 we find $R_0 = 21.9~\mu$m and $R_1 = 24.7~\mu$m. These radii are used in the BL and MCNP simulations.

\subsubsection{Particle Settling}\label{sssec:settling}
While the AGSF-33 particles are nominally matched with the density of water they, like most tracer particles, may have a non-uniform density within a batch. Density mismatches with the fluid induce relative motion due to buoyancy, which can bias velocity measurements. While for PIV the buoyancy induced velocity may often be neglected, the first flows measured with XPIV are likely to be creeping flows due to present technological limitations. Hence, consideration of settling speeds and buoyancy is warranted. When the relative velocity is low enough for Stokes drag to be appropriate (Re $<$ 1), the speed with which particles will rise or settle in a fluid can be calculated from equation \ref{eq:stokesDrag}.

\begin{equation} \label{eq:stokesDrag}
    U_{St} = d_p^2 \frac{\rho_{fluid} - \rho_{particle}}{18 \mu} g
\end{equation}
Here, $d_p$ is the particle diameter, $\rho_i$ is total density of $i$, $\mu$ is the dynamic viscosity of fluid, and $g$ is the acceleration due to gravity.

\begin{figure}
    \centering
    \includegraphics[width=0.5\textwidth]{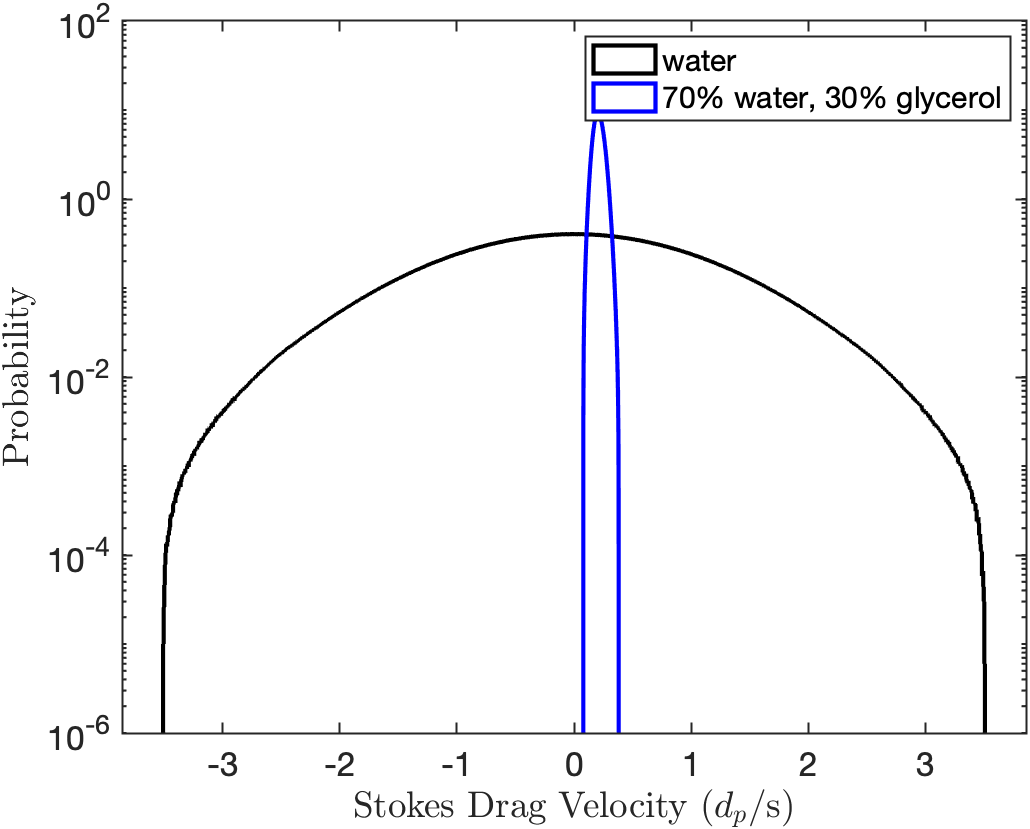}
    \caption{The probability density function of the terminal velocity (in particle diameters per second) for sieved, uniformly distributed, 45-53$\mu$m AGSF-33 tracer particles in the Stokes drag regime in water and in a 30\% glycerol, 70\% water mixture.}
    \label{fig:stokesMC}
\end{figure}

The manufacturer of AGSF-33 states that the particle densities can range from 0.9~g/cm$^3$ to 1.1~g/cm$^3$, with a nominal mean density of 1~g/cm$^3$. Additionally, we sieved the particles to between 45 and 53~$\mu$m. We assume that the particle diameters are distributed uniformly across the sieved range, and that the density is normally distributed with a 0.033~g/cm$^3$ standard deviation, truncated at 0.9~g/cm$^3$ and 1.1~g/cm$^3$. As a first approximation, we further assume that the diameter and density are independent random variables. 
Taking the Monte-Carlo approach, we simulate equation \ref{eq:stokesDrag} $10^8$ times for particles in water at 20$\degree$C, yielding the distribution shown in figure \ref{fig:stokesMC}. The distribution mean is $\leq$O($\pm 0.01~\mu$m/s); the standard deviation $\sigma= 49~\mu$m/s. As seen in figure \ref{fig:stokesMC}, 99.7\% of particles will travel less than three particle diameters in a second. Current XPIV limitations necessitate exposure times on the order of 10ms to 100ms. Over the course of a 100~ms exposure, 99.7\% of AGSF-33 particles in water will move less than 15~$\mu$m, or 0.3 particle diameters. It is desirable to minimize the particle motion in a single exposure. Mixing water and glycerol increases the fluid viscosity, thus dramatically reducing the particle Stokes velocity spread, as seen in figure \ref{fig:stokesMC}. This suggests that presently available particles may be suited to study a wide range of creeping flows with appropriately chosen fluid properties.

\section{X-Ray Image Simulation Methodology} \label{sec:sims}
\subsection{Beer-Lambert Simulation}\label{ssec:BL}
\begin{figure}
    \centering
    \includegraphics[width=0.5\textwidth]{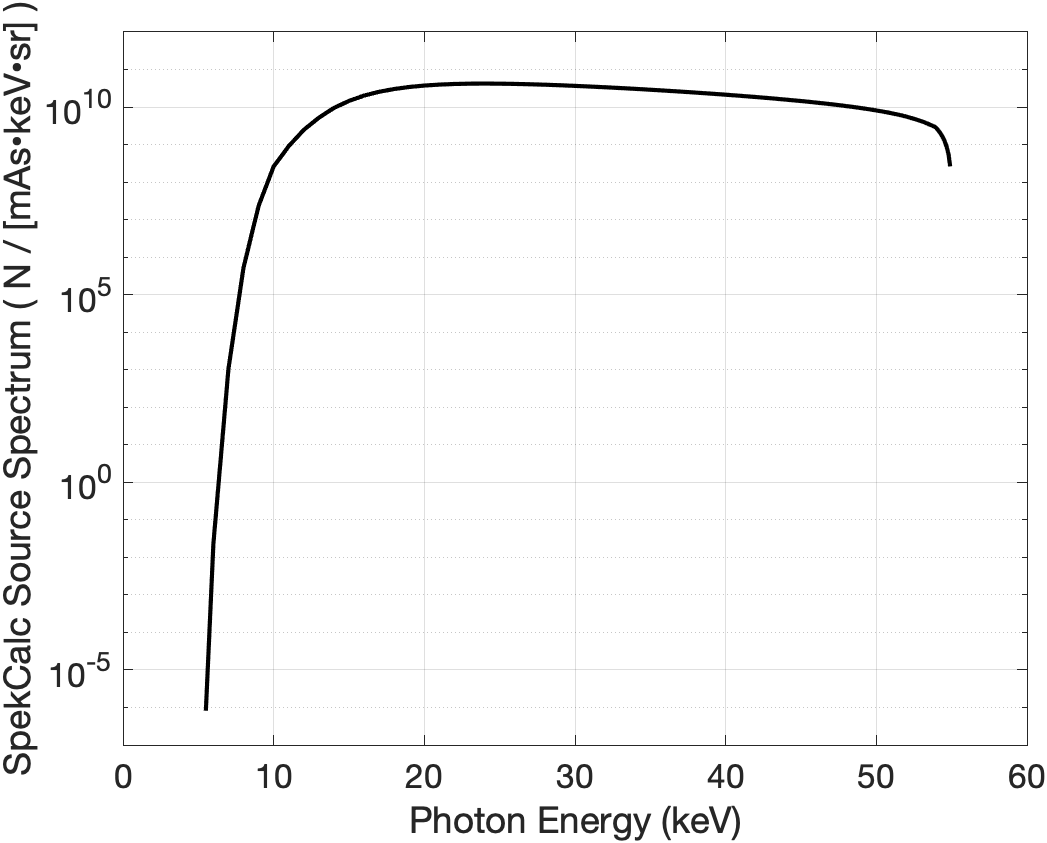}
    \caption{SpekCalc source spectrum of a 22.5$\degree$ half-angle conical tungsten target at 55~kV with a 1~mm thick aluminum window.}
    \label{fig:sourceSpec}
\end{figure}

Equation \ref{eq:BL} is the Beer-Lambert law, which describes the attenuation of photons at energies $\epsilon$ as they pass through $J$ media of thickness $s_j$ that have mass attenuation coefficient $\mu_j(\epsilon)$ and density $\rho_j$. The attenuation coefficient for each material is obtained from the NIST X-ray attenuation database \cite{j_h_hubbell_x-ray_2004}.
\begin{equation} \label{eq:BL}
    I = \int_{E_{thresh}}^{E} Q(\epsilon) I_0(\epsilon) \Delta t e^{ -\sum_{j=1}^{J}{ \rho_j \mu_j(\epsilon) s_j } } d\epsilon
\end{equation}
Here, $E_{thresh}$ is the detector threshold energy, $E$ is the maximum photon energy, $Q$ is the efficiency of the detector, $\Delta t$ is the exposure time, $I$ is the detected photon intensity, and $I_0$ is the source photon intensity, the spectrum for which can be seen in figure  \ref{fig:sourceSpec}.

\begin{figure}
\centering
\includegraphics[width=0.5\textwidth]{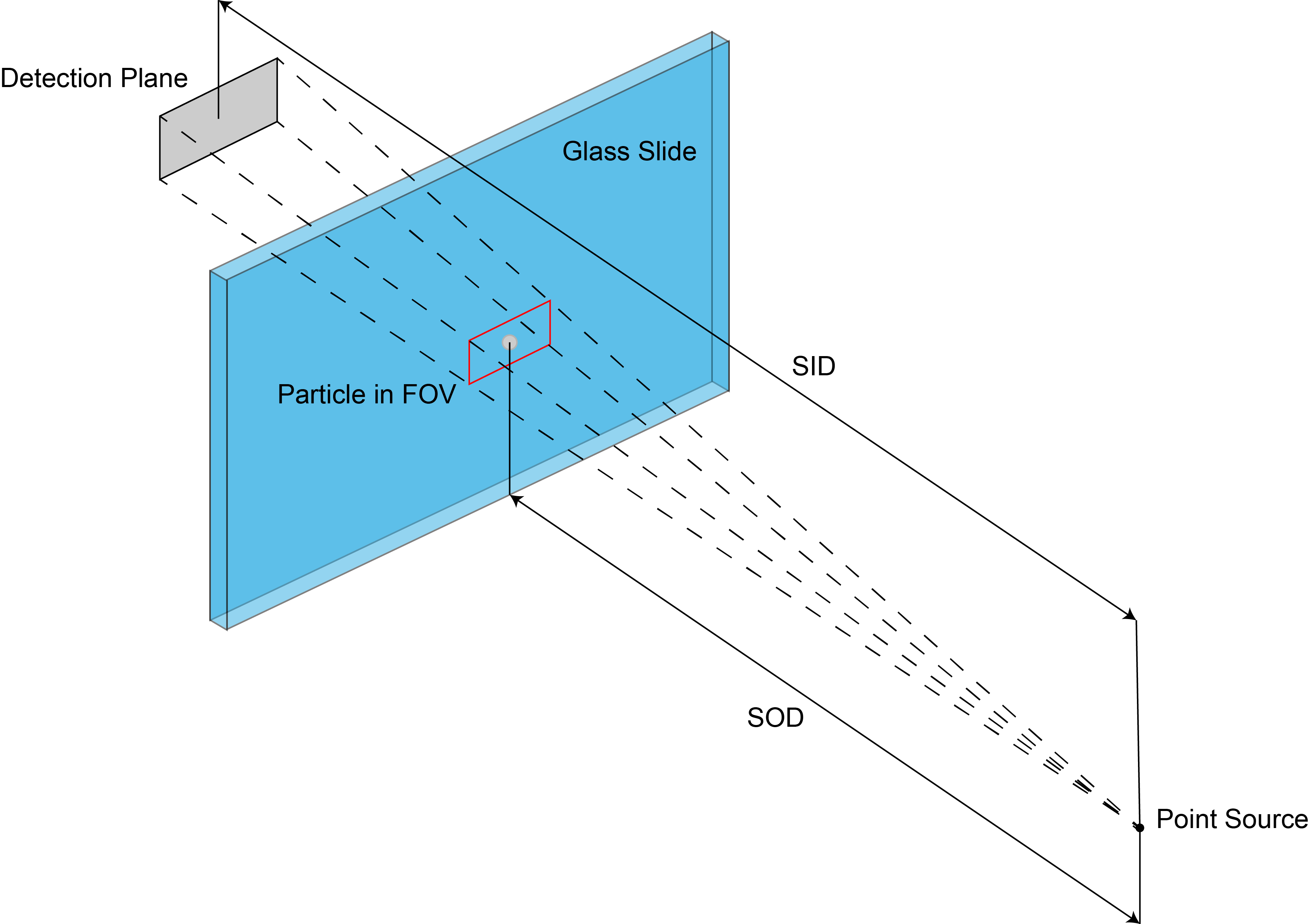}
\caption{A point source generates rays along which the Beer-Lambert law is evaluated at the detector plane. The detector itself is not in the geometry. Here, an AGSF-33 particle is placed on the front surface (surface towards source) of a glass slide. Figure not to scale.}
\label{fig:BLgeom}
\end{figure}

Equation \ref{eq:BL} is a steady state form of the Beer-Lambert equation. Assuming the subject being imaged does not move or change in composition over the course of a single exposure (PIV exposures are ideally shorter than the characteristic time of a flow) the steady state form can predict a detector image. A diagram of the BL simulation geometry as implemented in the present study can be seen in figure \ref{fig:BLgeom}. The SID is 47.6~cm as measured from the center of the detector. An AGSF-33 particle is set on a glass slide and the SOD as measured from the center of the particle to the source was 1~cm. The AGSF-33 particle geometry is discussed in Section \ref{ssec:particleChoice}.

A ray-tracing MATLAB code was developed to calculate the number of photons that are detected by each pixel. Critically, each pixel is divided into sub-pixels to account for geometry projection variations across a single pixel. The error associated with sub-pixel variation was found to be negligible (below 10\% in the worst case scenario) when 121 (11$\times$11) sub-pixels were considered. The Dectris Pilatus3 crystal surface components are to the first order transparent to the incoming radiation, so the pixels have a 100\% fill factor. We therefore consider each sub-pixel to detect photons equally. The convergence of the pixel photon count with increasing sub-pixels is discussed in Appendix \ref{app:A}. The photon intensity for each sub-pixel is calculated using equation \ref{eq:BL}. For each object along the ray, the thickness $s_j$ was calculated by taking the Eulerian distance between the two intersection points (entry and exit) of the ray with the object. The detected photon intensity for all sub-pixels within a pixel are summed to calculate the number of detected photons for that pixel. In this way, the number of detected photons are calculated for each pixel in the detector, thus initially simulating a noiseless X-ray image taken with a PCD.

While PCDs exhibit no Gaussian thermal white noise, the emission process of the X-ray source and the binomial nature of photon attenuation introduce Poisson noise. To simulate a more realistic image, Poisson noise is added to the image calculated with equation \ref{eq:BL}. To add Poisson noise, we take each noiseless calculated pixel value to be the expected value of a Poisson distribution for that pixel. A random value is selected from said distribution to be the noisy pixel value. It is important to note that adding pure Poisson noise assumes that the effects of photon pileup are negligible \cite{wang_pulse_2011}.

The source intensity spectrum $I_0$, shown in figure \ref{fig:sourceSpec}, is calculated using SpekCalc \cite{poludniowski_calculation_2007, poludniowski_calculation_2007-1, poludniowski_spekcalc_2009}. The true source intensity spectrum and spatial distribution are not known precisely. In order to calculate the number of photons to simulate for a given exposure time, experimental flat images are used to calculate the average number of photons emitted from the source that were detected. The solid angle of the detector is known; the source has a 15$\degree$ cone angle, so the total number of source photons is calculable, assuming uniform emission. Attenuation through 47.6~cm of air was accounted for by taking the average air attenuation ratio across all photon energies, then calculating an estimate for $I_0$ based on equation \ref{eq:BL}. In reality, the source emission is not spatially uniform due to the heel effect, and the attenuation of X-rays through air is strongly energy-dependent. Improved source characterization could eliminate errors originating from spectrum and flux approximations, improving simulation accuracy. In the absence of an X-ray source and detector to measure the source flux, SpekCalc has been found to adequately estimate the source flux in addition to the intensity spectrum. In this study, we approximately measure flux as described above, but rely on SpekCalc for the intensity spectrum.

While BL simulations are computationally cheap, the approach entails several simplifying assumptions. Most consequential of these is that BL simulations assume all scattered photons are attenuated and not detected. In general, this is not true. Many scattered photons may experience a shallow scattering angle and will be detected even if they are scattered from their original ray. Photons may also experience scatter within the detector, a process discussed in greater detail in section \ref{ssec:mcnpProcess}. As a result of the scattering attenuation assumption, BL will underestimate the number of detected photons and noise intensity.

For computational efficiency, source photons are emitted uniformly from a single, infinitesimal point. We assume uniform emission for simplicity, but in reality the source emission is not uniform due to the heel effect. An infinitesimal source neglects the blurring effects from a finite focal spot. Blurring from a finite focal spot reduces image contrast. One could account for a finite focal spot in BL simulations with distributed emission locations, spatial non-uniformity of emissions, and randomness of emission flux could additionally be considered. However, the computational advantage of BL simulations as a design tool would diminish if we consider thousands of emission sources and hundreds of sub-pixels.

Although we make simplifying assumptions to reduce the computational cost of the BL simulations, the MATLAB code developed for this study is not optimized for computational efficiency. The calculations are done on a CPU and are not parallelized. Parallelization on a GPU is low-hanging fruit for improvement. There is a wealth of literature on ray-tracing methods \cite{glassner_space_1984, badouel_distributing_1994, garanzha_fast_2010, vidal_accelerated_nodate} that the reader can refer to for pathways to speed up the BL simulation method. For more recent advances in ray-tracing, we refer the reader to \cite{wikipedia_path_2021, mcguire_introduction_nodate}. For the purposes of this study, we simply consider the accuracy of the BL simulations compared to real X-ray images as the benchmark of simulation performance as a potential design tool.

Lastly, X-ray fluorescence \cite{david_attwood_x-rays_nodate} is not considered in the BL simulations. We ran identical simulations using MCNP to observe the effects of scattering, fluorescence, and a finite focal spot. MCNP is discussed in section \ref{ssec:mcnp}.

\subsection{Monte Carlo N-Particle Simulations}\label{ssec:mcnp}
Monte Carlo N-Particle (MCNP) is a particle transport simulation code developed by Los Alamos National Lab \cite{cj_werner_et_al_mcnp62_2018}. In MCNP, each photon is assigned an energy, departure vector, and position in the focal spot based on known or assumed distributions. Using databases compiled from experiment data over the past five decades, MCNP calculates the probability of different photon interactions with a medium as well as the probability of the outcome of said interactions. For example, as a photon passes through water, there are probabilities for that photon being scattered, absorbed, or passing through unperturbed. By simulating as many photons as the experiment source emits for a given exposure time, one can reconstruct the detector image. Although MCNP is more computationally expensive than the BL simulations discussed in \ref{ssec:BL}, it incorporates scattering effects both in the domain and the detector, a finite source size, and fluorescence.

MCNP is a highly optimized program that has been used, particularly in the nuclear engineering and radiation protection communities, for nearly five decades. The user provides the MCNP program with an input file that dictates the domain, geometry, source characteristics, data to be exported etc., without modifying the MCNP code itself.

\subsubsection{MCNP Simulation Setup}
\begin{figure}
    \centering
    \includegraphics[width=0.5\textwidth]{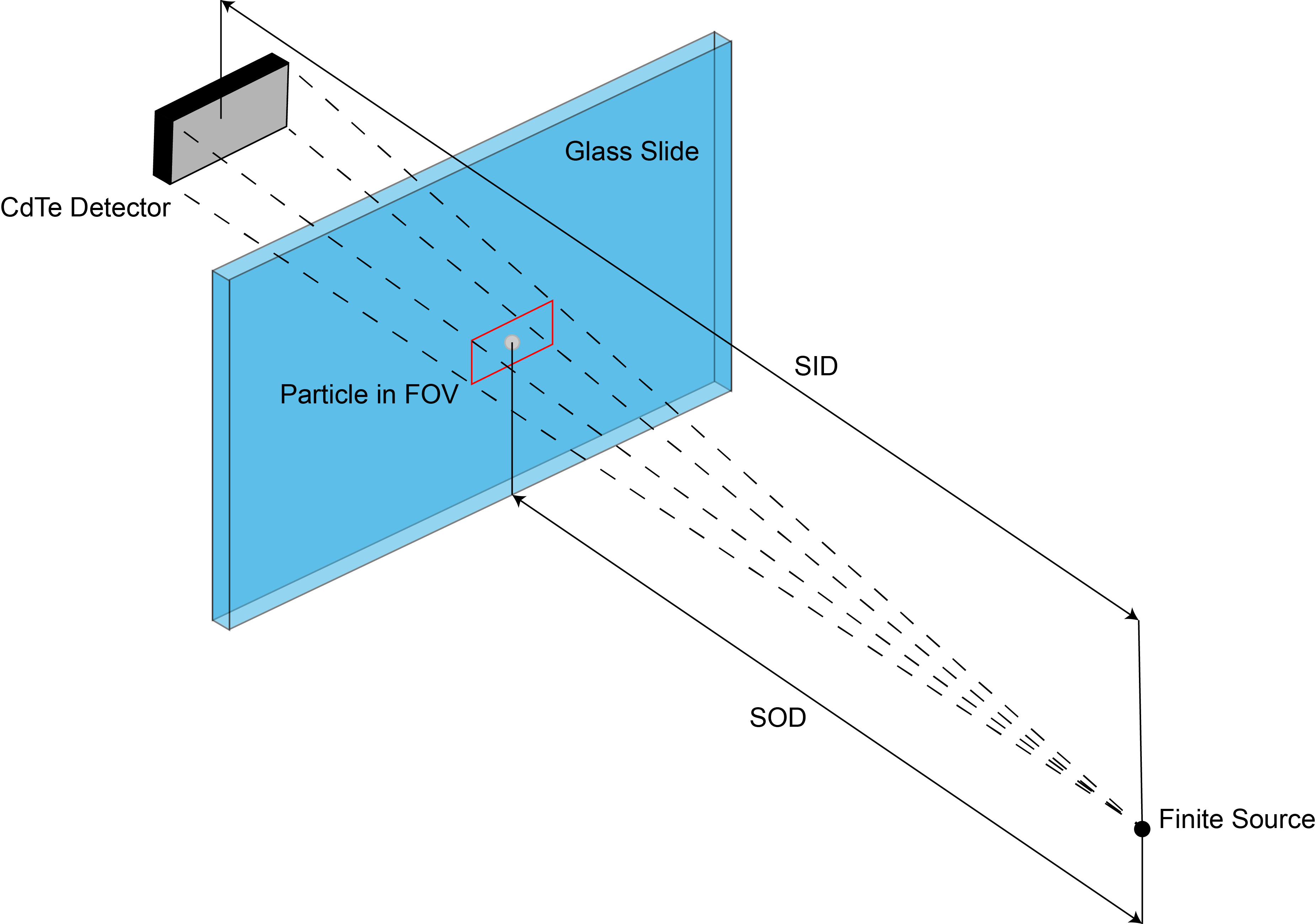}
    \caption{The MCNP geometry is similar to the BL simulation setup of figure \ref{fig:BLgeom}, but includes a CdTe-detector crystal of finite thickness (omitting housing and electronics for simplicity) and accounts for the finite X-ray source spot. Figure not to scale.}
    \label{fig:MCNPgeom}
\end{figure}

The MCNP domain geometry defined for the present study can be seen in figure \ref{fig:MCNPgeom}. A 6~$\mu$m radius finite source emits a 15$\degree$ conical beam with uniform photon flux \cite{noauthor_yxlon_nodate}. The source photon energy probability density function is calculated from the same SpekCalc spectrum used for the BL simulation. Unlike the BL simulations, which do not have a detector in the domain, the detector in MCNP is modeled as a uniform CdTe panel in the domain. In order to reduce unnecessary photon tracking, the domain is 60~cm by 20~cm by 20~cm with non-reflective boundary conditions. In other words, it is assumed that any photons that leave this domain will not scatter back to the detector.

MCNP simulates a path for each source photon with the Monte Carlo method based on the probabilities of an interaction between the photon and the domain object(s). If a photon enters the CdTe panel, the location and energy of that photon are outputted as the photon scatters or is absorbed and deposits energy within the CdTe panel. Then, in post-processing, we count how many photons entered each pixel and deposited enough energy to be detected, i.e., deposited energy greater than or equal to our experiment energy threshold. In a PCD such as the one we use in this study, the simulated image pixel intensity is simply the number of photons that are counted for each pixel. The post-processing to generate the image is done in MATLAB.

Notably, MCNP does not simulate the number of source photons emitted per unit time, but rather simulates a fixed number of photons that in the physical world corresponds to an actual emission time with some probability. Using the same procedure as was described in \ref{ssec:BL}, MCNP simulated approximately the same number of source photons as the BL simulations and experiment images, thus approximating an image with an exposure time equal to the experiment images.

\subsubsection{MCNP Processing Methodology}\label{ssec:mcnpProcess}
The output method of choice for our purposes is the particle track (PTRAC) file. A PTRAC file contains every interaction that every photon has with the environment. To simulate the image formed at the detector, two filters are applied to the PTRAC file in order to restrict the output to relevant interactions. The first filter reduces the reported interactions to only those that occur within the detector. The second filter is applied to only include interactions that would generate a signal within the photon-counting detector. A snippet of a PTRAC file can be seen in figure \ref{fig:ptrac} in Appendix \ref{app:B} along with a detailed explanation. A PTRAC file output gives us the ability to change pixel size and detector energy thresholds a posteriori. Since MCNP simulations take so long to run, it is beneficial to be able to reuse one set of output data. As mentioned previously, the PTRAC file(s) are inputted into MATLAB for post-processing to generate the simulated image.

\begin{figure}[h]
    \centering
    \includegraphics[width=0.5\textwidth]{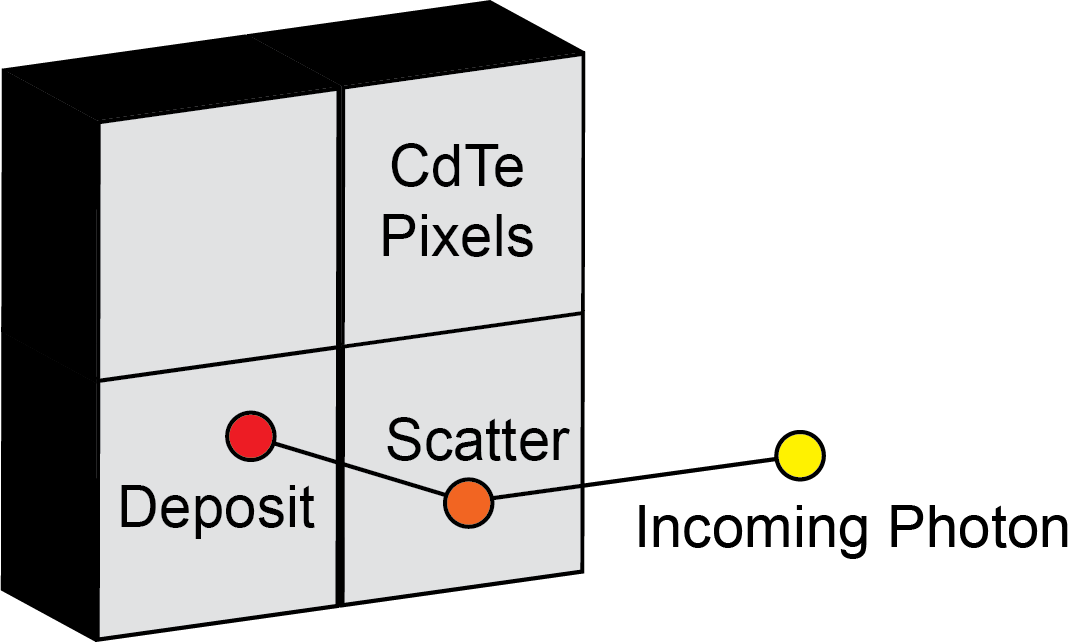}
    \caption{An incoming photon can enter one pixel, deposit sufficient energy to be counted, then scatter to another pixel where it is counted once more.}
    \label{fig:intScatter}
\end{figure}

\begin{figure}[H]
    \centering
    \includegraphics[width=0.45\textwidth]{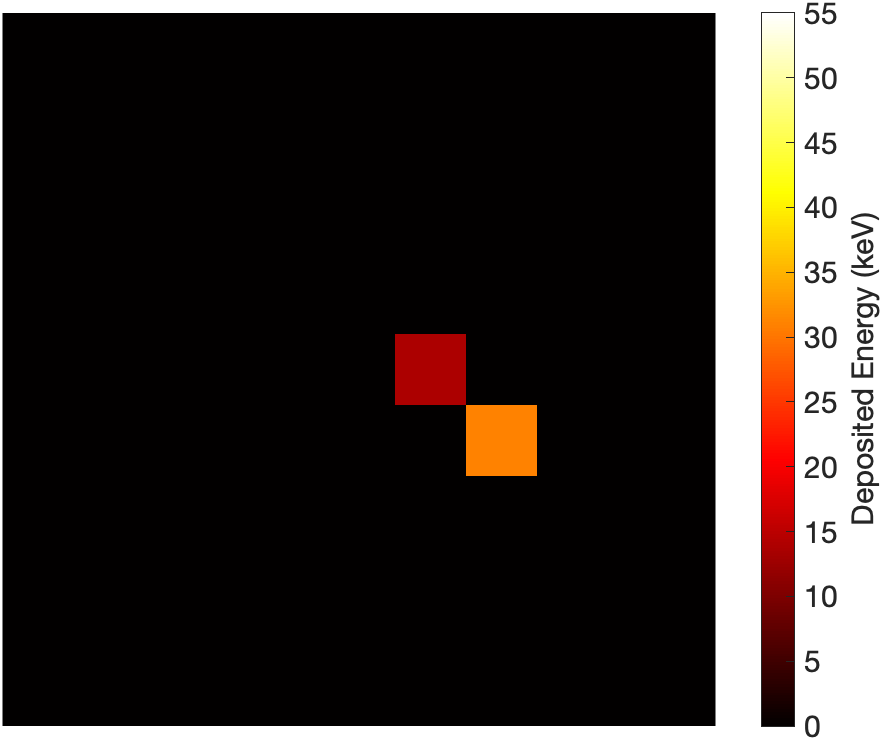}
    \caption{A single photon in an MCNP simulation is detected in two separate pixels. Multi-detections such as this one lower the image SNR.}
    \label{fig:multiDet}
\end{figure}

For the present study, a major benefit of using a PTRAC file output in combination with a CdTe panel in the domain is the ability to simulate multi-detection photons. Multi-detection photons are photons that deposit sufficient energy to be detected in one pixel before scattering to an adjacent pixel and being detected again, a process we refer to as internal scattering. A single photon can then be detected two or more times. Figure \ref{fig:intScatter} depicts how multi-detection could occur. Figure \ref{fig:multiDet} shows a multi-detection event from an MCNP simulation.

In order to ascertain the impact of internal scattering, one set of MCNP simulation data was processed in two ways. First, the simulation data was processed such that a photon was detected once in the pixel it first enters. Processing in this way will not count multi-detections. Next, the simulation data was processed to include internal scattering effects. The difference between the former and latter processing is approximately 8\% of the average multi-detection pixel count. Internal scattering clearly should be included when simulating a PCD image. The ability to do so in MCNP is an advantage over BL.

\section{Results and Discussion} \label{sec:res&dis}
\subsection{Experiment and Simulation Image Comparison} \label{ssec:slideResults}
\begin{figure}[H]
    \centering
    \begin{subfigure}[b]{0.45\textwidth}
        \includegraphics[width=\textwidth]{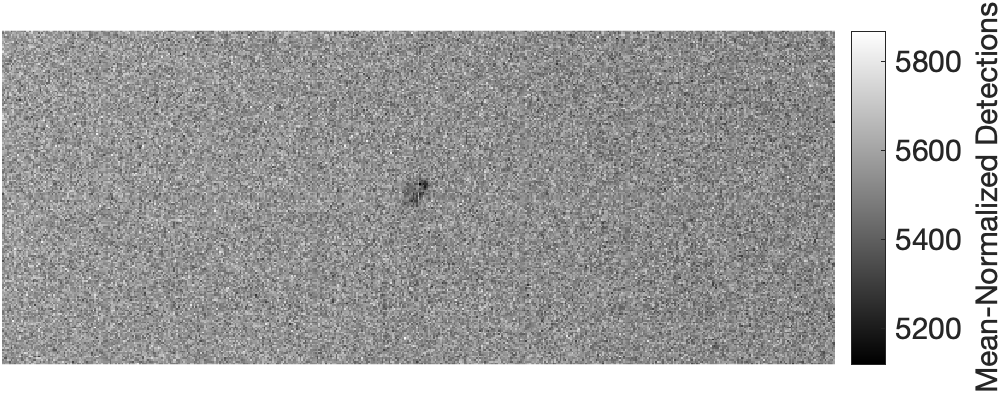}
        \caption{Flat field corrected X-ray image of an AGSF-33 particle on a glass slide. The particle can be seen in the center of the image.}
        \label{fig:expImg}
    \end{subfigure}
    \hfil
    \begin{subfigure}[b]{0.45\textwidth}
        \includegraphics[width=\textwidth]{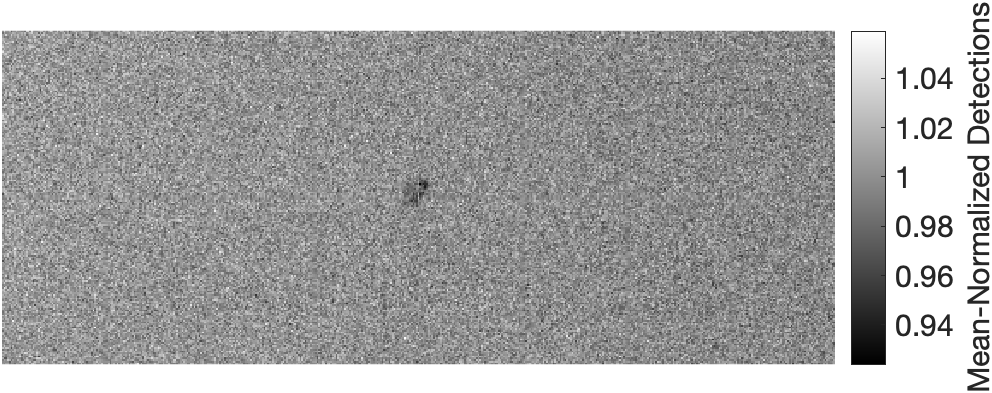}
        \caption{The mean-normalized X-ray image of data shown in (a). The average pixel intensity is 5,540 photons.}
        \label{fig:expImgnorm}
    \end{subfigure}
    \caption{A sample image from the experiment images collected at 55~kV, 500~$\mu$A for a 100~ms exposure time.}
    \label{fig:slideImgExp}
\end{figure}

\begin{figure}[H]
    \centering
    \includegraphics[width=0.5\textwidth]{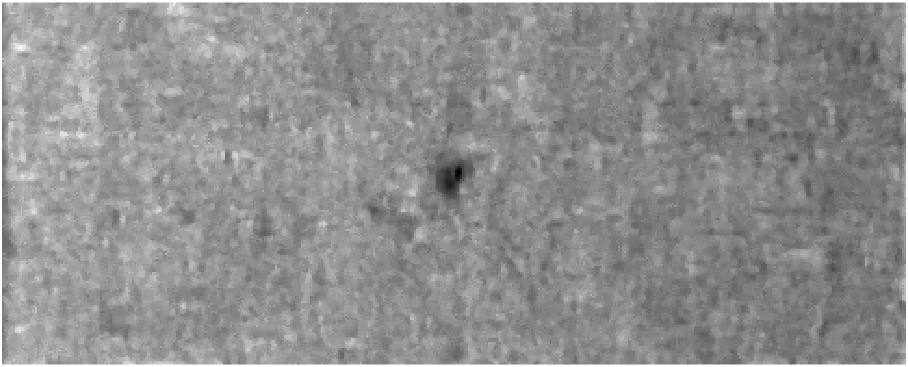}
    \caption{A PURE-LET deconvolution then a CLAHE filter with a square span of 8$\times$8 pixels is applied to the glass slide image in figure \ref{fig:slideImgExp}. A median filter with a 3$\times$3 span is applied last. The particle contrast improves, which in turn would improve the PIV correlation peak.}
    \label{fig:slideExpfilt}
\end{figure}

Figure \ref{fig:slideImgExp} shows a flat field corrected image of a particle taken at a source voltage of 55~kV, target current of 500~$\mu$A, and a 100~ms exposure time. The nominal detector threshold is 20~keV. The SOD is 1~cm; the SID is 47.6~cm. In figure \ref{fig:slideImgExp} we see a single AGSF-33 particle taped to a 1~mm thick glass slide. Although the particle is clearly detectable, the ability to localize the particle for PIV can be enhanced with simple de-noising techniques specific to Poisson-noisy images. Figure \ref{fig:slideExpfilt} shows the image in figure \ref{fig:slideImgExp} after applying a PURE-LET deconvolution \cite{li_pure-let_2018}, a CLAHE filter \cite{pizer_adaptive_1987}, and finally a median filter. After filtration, the particle is even more visible. PIV image correlation could be performed on an image containing many of these particles. For the purposes of this paper, however, we will not attempt to optimize settings for localizing a single particle in the shortest possible exposure. Instead, we will focus our discussion on the simulation accuracy and design utility.

\begin{figure}[H]
    \centering
    \includegraphics[width=0.45\textwidth]{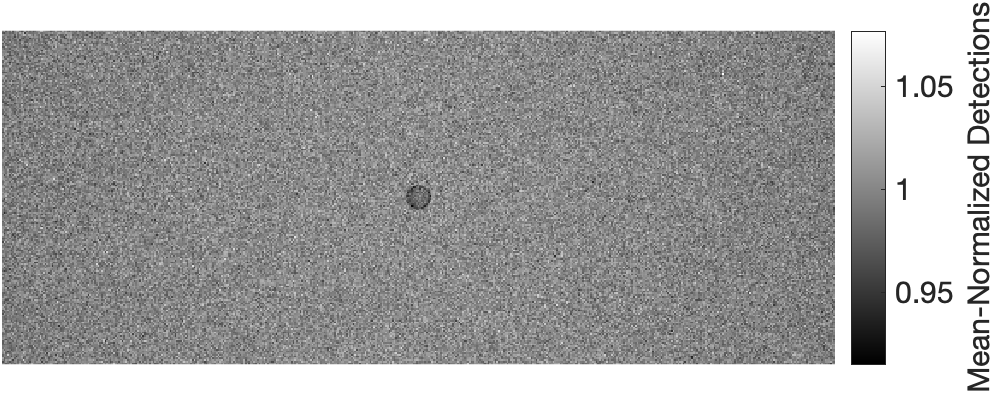}
    \caption{A normalized BL simulated Poisson-noisy image. The simulated source flux was matched to the experimentally calculated value. The average pixel intensity is 4,661 photons.}
    \label{fig:BLnoise}
\end{figure}

\begin{figure}
    \centering
    \begin{subfigure}[b]{0.45\textwidth}
        \includegraphics[width=\textwidth]{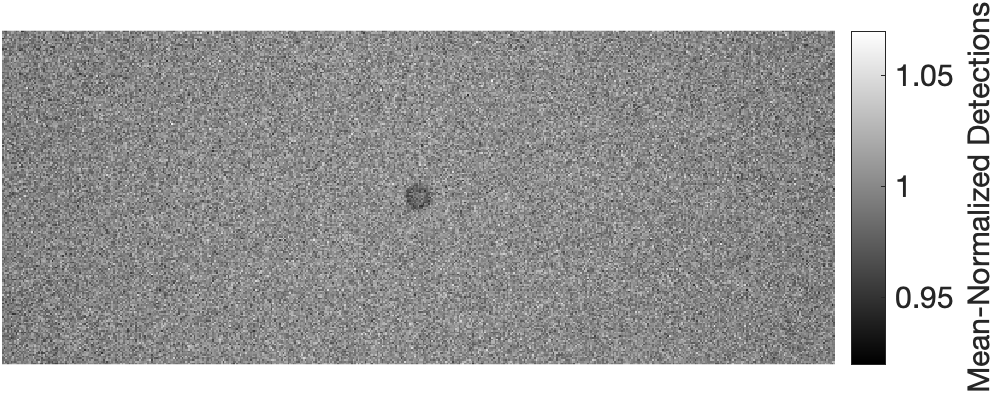}
        \caption{55~kV, 500~$\mu$A BL simulation with Gaussian blur; $\sigma=1$, blur span $= 5$ pixels square.}
        \label{fig:gblur1}
    \end{subfigure}
    \vfill
    \begin{subfigure}[b]{0.45\textwidth}
        \includegraphics[width=\textwidth]{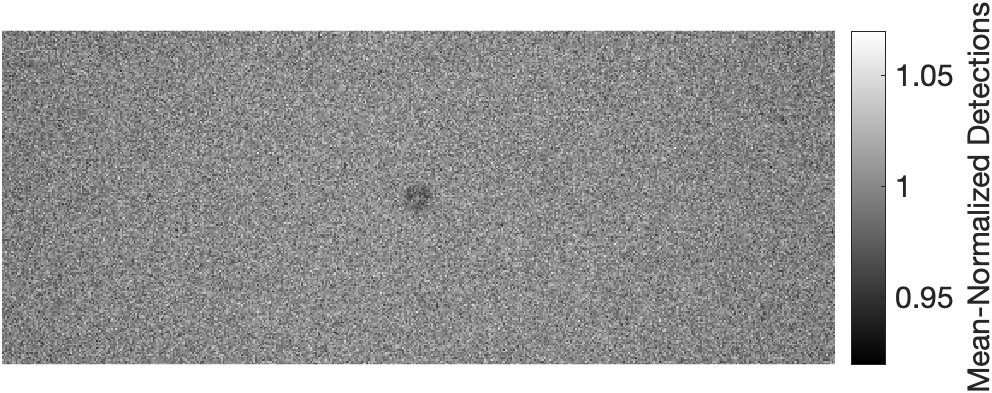}
        \caption{55~kV, 500~$\mu$A BL simulation with Gaussian blur; $\sigma=2$, blur span $= 9$ pixels square.}
        \label{fig:gblur2}
    \end{subfigure}
    \vfill
    \begin{subfigure}[b]{0.45\textwidth}
        \includegraphics[width=\textwidth]{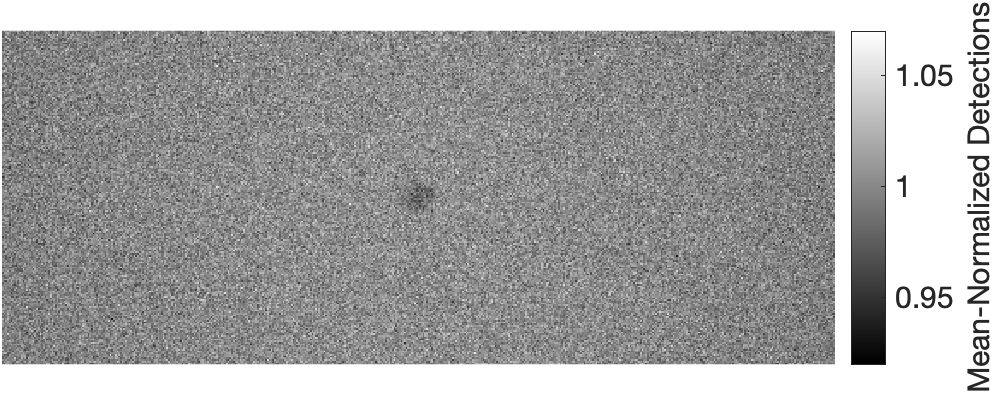}
        \caption{55~kV, 500~$\mu$A BL simulation with Gaussian blur; $\sigma=3$, blur span $= 13$ pixels square.}
        \label{fig:gblur3}
    \end{subfigure}
    \caption{Gaussian blur degrades the SNR of the BL simulated image. In the future, a finite source focal spot may be simulated by adding Gaussian blur. The average pixel intensity for all images is 4,661 photons.}
    \label{fig:gblur}
\end{figure}

Figure \ref{fig:BLnoise} shows the BL image with Poisson noise added. The simulation settings are the same as the experimental settings for figures \ref{fig:slideImgExp} and \ref{fig:slideExpfilt}, with the source flux matched. The BL simulation assumes a point source, which produces an image without the realistic blurring due to a finite focal spot. In the BL image, the silver coating is clearly resolved, whereas in the experiment image the coating is blurred with the rest of the particle. Simple blurring can be added to the BL image without the computational cost of simulating multiple source locations if we add Gaussian blur, as seen in figure \ref{fig:gblur}. Clearly, blurring degrades SNR; figure \ref{fig:gblur3} qualitatively resembles the experiment data. As previously noted in section \ref{ssec:BL}, more could be done to quantitatively address image blurring in BL simulations while balancing computational expense.

The most significant limitation of the BL simulation, however, is that it cannot simulate internal or external scattering. As noted in the caption of the normalized images, the average intensity of the BL image is notably lower than the experiment data. The depressed photon detection count can be attributed to the attenuation assumption for scattered photons. Scattered photons in BL are not detected, which in general is not true. Many scattered photons may experience shallow scattering angles, and so could still be detected. Furthermore, photons may be detected multiple times due to scattering within the detector panel.

\begin{figure}
    \centering
    \includegraphics[width=0.5\textwidth]{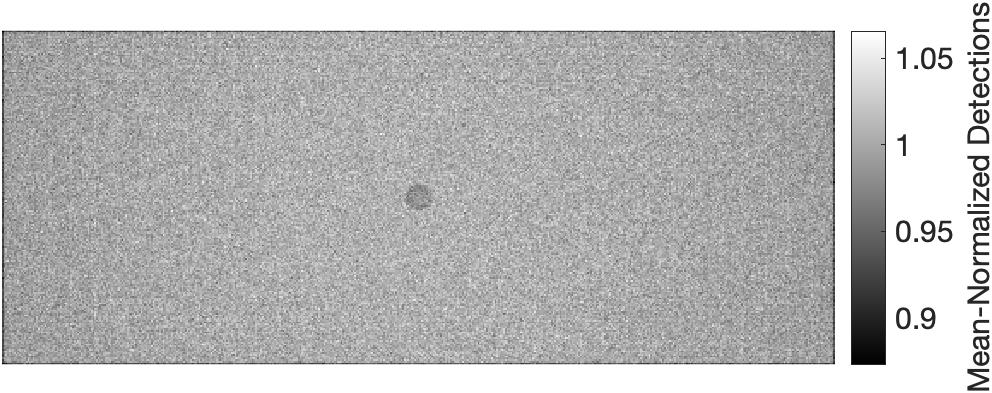}
    \caption{The MCNP simulation image of an AGSF-33 particle on a glass slide. The MCNP image shows blurring from the finite focal spot. The average pixel intensity is 5,074 photons.}
    \label{fig:MCNPslide}
\end{figure}

The MCNP image, seen in figure \ref{fig:MCNPslide}, matches the experiment image better than the BL image seen in figure \ref{fig:BLnoise}. Qualitatively, the first thing we notice is that the particle is slightly blurred, primarily due to MCNP modeling an idealized circular finite focal spot. If all blurring were attributable to focal spot blurring, the blur kernel should be approximately 1.6 pixels across. The MCNP image qualitatively matches figure \ref{fig:gblur1}, which has a kernel standard deviation of $\sigma = 1$ pixel. However, the experiment image qualitatively agrees best with figure \ref{fig:gblur3},  where the kernel standard deviation is $\sigma=3$ pixels. This suggests that additional blur is attributable to a non-circular focal spot, the resulting heel effect, detector imperfections, or some combination thereof.

\begin{figure}
    \centering
    \includegraphics[width=0.5\textwidth]{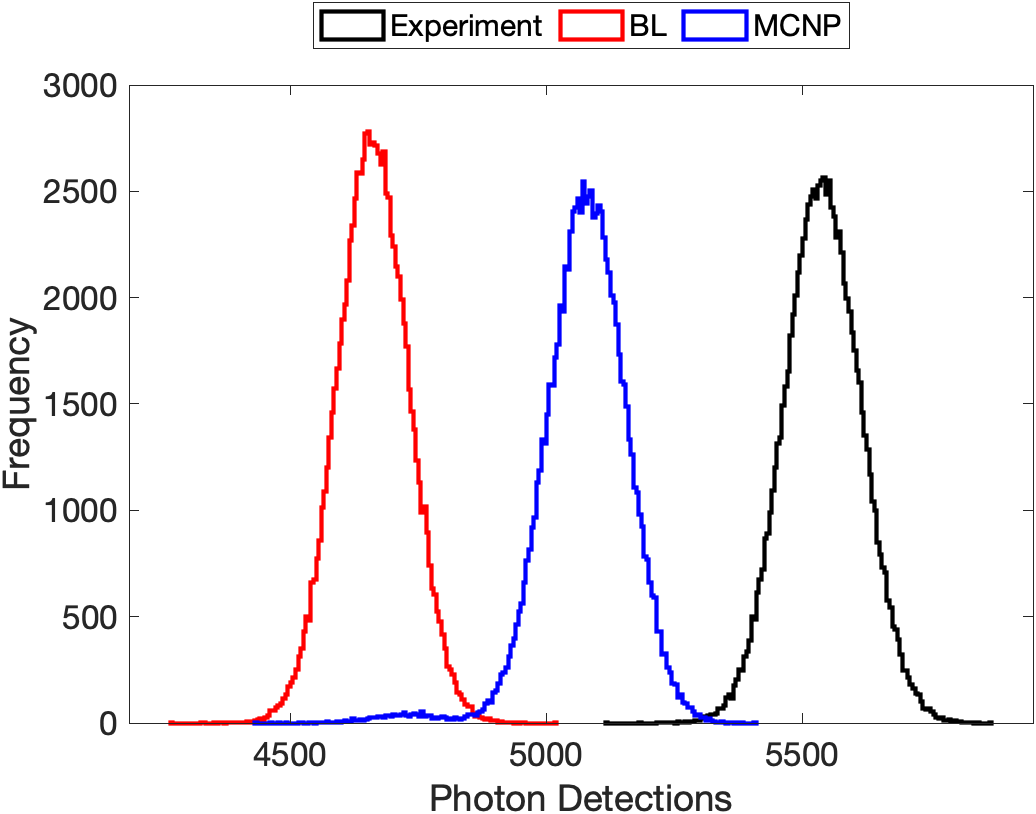}
    \caption{The MCNP, BL, and experiment image photon detection distributions. A small peak to the left in the MCNP distribution is due to the depressed photon counts in the MCNP edge pixels. MCNP is more accurate than BL, but still does not match experiment exactly.}
    \label{fig:mcnpExphist}
\end{figure}

Quantitatively, the average photon detections per pixel in the MCNP image approaches that of the experiment image, suggesting that scatter attenuation is the main reason for depressed photon detections in BL. Figure \ref{fig:mcnpExphist} compares the distribution of photon detections per pixel from the experiment, BL, and MCNP. The distribution moment data are shown in table \ref{tab:distMom}. While MCNP more closely predicts the experimental statistics, it still does not align exactly with what was measured. This is to be expected, since MCNP cannot account for detector crystal imperfections, anisotropy in the detector, or non-Poisson noise behavior due to photon pile-up in the detector. MCNP did not simulate photons scattering to the detector from outside the domain or the detector housing and electronics, as these external materials were not included in the simulation geometry. Furthermore, neither MCNP nor BL can account for inaccuracies in the simulated source spectrum or calculated source flux. Improving the source characterization is one way to improve the simulation accuracy in the future.

\begin{table}[]
    \centering
    \begin{tabular}{|c|c|c|}
        \hline
         & Mean & Variance \\
        \hline
        \hline
        Experiment & 5,540 & 5,622 \\
        \hline
        MCNP (with edges) & 5,074 & 7,344 \\
        \hline
        MCNP (without edge pixels) & 5,079 & 5,604 \\
        \hline
        BL & 4,661 & 4,803 \\
        \hline
    \end{tabular}
    \caption{BL, MCNP, and experiment image photon detection distribution moments.}
    \label{tab:distMom}
\end{table}

In figure \ref{fig:mcnpExphist} there is a slight bump in the MCNP histogram at about 4,800 photon detections. This bump is an artifact of the MCNP model geometry. Our model geometry does not have any material surrounding the CdTe panel, so the edge pixels have less adjacent material from which photons may scatter back into the detector pixels. Statistics in table \ref{tab:distMom} are provided for the MCNP image with and without the edge pixels.

Despite the greater computational expense, it may be best to use MCNP for imaging conditions in which scatter, fluorescence, and focal spot blurring are more substantial. In the following section, however, we demonstrate that in some cases, BL can offer effectively the same accuracy as MCNP at a fraction of the computational cost.

\subsection{Predicting Image SNR For XPIV System Design}
\begin{figure}
    \centering
    \includegraphics[width=0.5\textwidth]{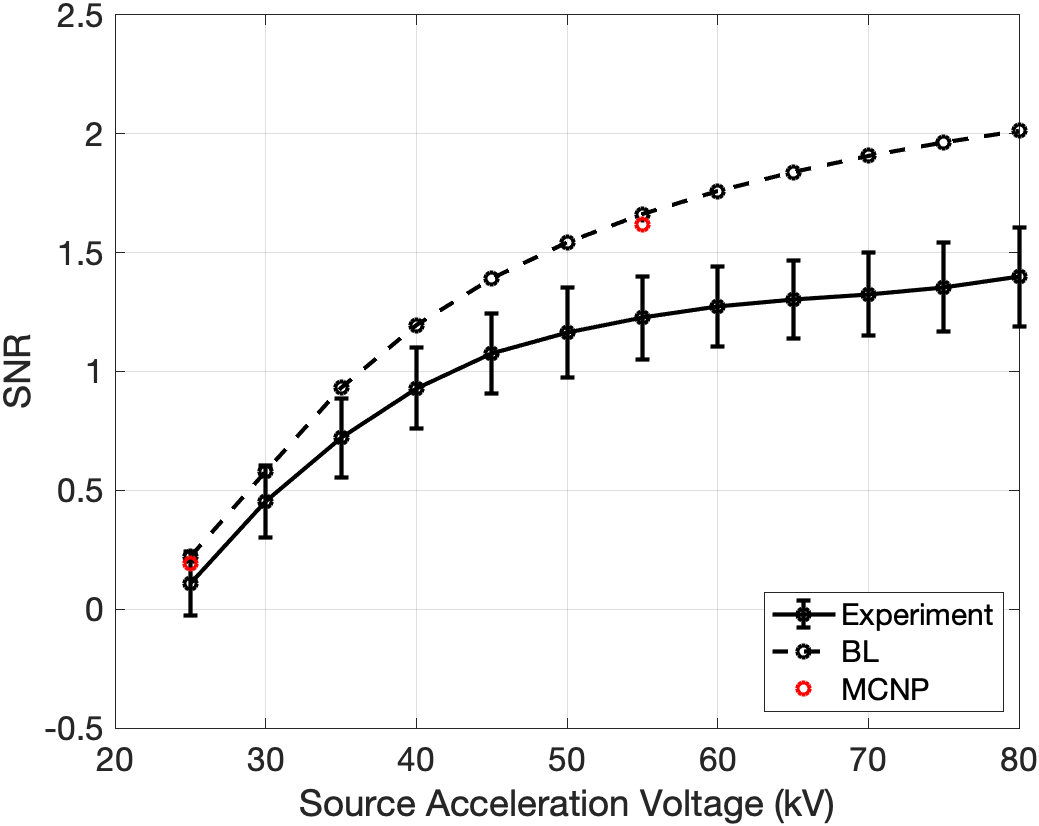}
    \caption{The image SNR from experiments and BL simulations as a function of the source voltage. The MCNP SNR for the 55~kV image is shown for comparison.}
    \label{fig:SNR_kV}
\end{figure}

MCNP and BL simulations can be used as tools to help XPIV experiment designers methodically improve their system. BL simulations prove useful in design applications such as predicting image SNR behavior. SNR can be used as a metric to gauge the detectability of tracer particles in an image. XPIV system performance improves as the particle detectability improves. Figure \ref{fig:SNR_kV} shows the SNR as a function of source voltage for the experimental, BL, and MCNP images. As expected, the BL simulations overestimate the SNR; MCNP is more accurate, but only slightly. Practically speaking, BL simulations are the preferred predictive tool by virtue of being computationally inexpensive. As figure \ref{fig:SNR_kV} demonstrates, BL simulations are able to explore a wide parameter space relatively quickly, unlike MCNP.

The SNR is calculated as the ratio of the particle-background contrast and the background noise intensity. The particle contrast was calculated by taking the difference in average pixel value for the particle and for all pixels within one particle diameter of the particle edge (i.e., the background pixels). The contrast was then divided by the standard deviation of the background pixels. The standard deviation approximates the background noise intensity. For the BL SNR, the standard deviation was calculated assuming a perfect Poisson distribution, so the standard deviation is taken to be the square root of the average background pixel value.

\begin{figure}
    \centering
    \includegraphics[width=0.5\textwidth]{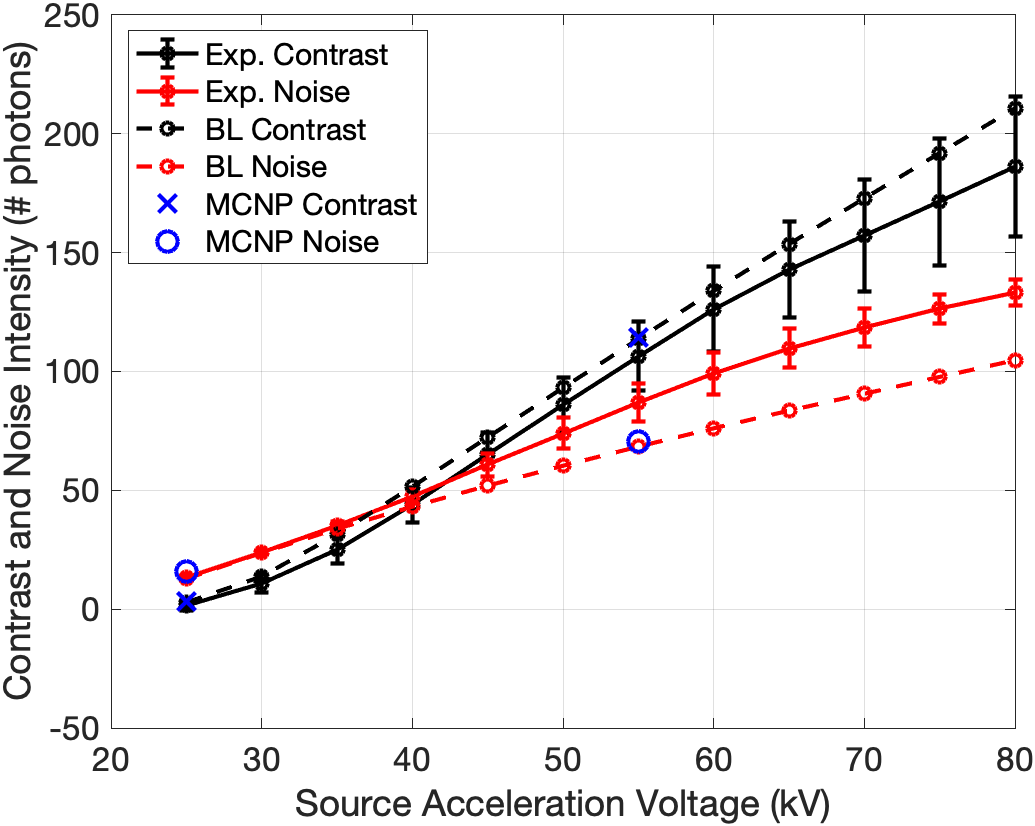}
    \caption{The image contrast and noise intensity are plotted as a function of the source voltage. BL underestimates the noise intensity, likely due to the simplistic scattering model.}
    \label{fig:cont_noise}
\end{figure}

The BL simulations predict the SNR behavior best in the low source voltage regime where scattering is less likely and the Poisson noise regime holds. However, the predicted and measured curves diverge at higher source voltages as BL underestimates the noise intensity and overestimates the contrast, as seen in figure \ref{fig:cont_noise}. The contrast is higher in the BL images because there are no blurring effects, as discussed in figure \ref{fig:gblur}. MCNP, on the other hand, incorporates blurring effects due to scatter and a finite focal spot with ease, and as a result predicts contrast slightly more accurately. Both MCNP and BL underestimate the noise intensity. This makes sense, considering both BL and MCNP are limited to Poisson noise and idealize the detector. Experiment images must contend with detector artifacts and anisotropy in the CdTe crystal. The noise intensity is also a function of the number of source photons, so improving source flux modeling is crucial for improving SNR prediction.

It is important to note that in PCDs in particular the image noise is not perfectly Poisson due to pileup effects \cite{wang_pulse_2011}. In some situations it may be advantageous to consider energy integrating or scintillator-based X-ray imagers. When the photon flux rate is small compared to the maximum photon count rate of the imager, the effects of pileup can be neglected and the noise can be taken to be Poisson. The BL and MCNP simulations neglect pileup effects. As the source voltage increases, modeling image noise as Poisson becomes less tenable, which could partially explain the noise intensity divergence seen in figure \ref{fig:cont_noise}, in addition to the effects mentioned previously.

\subsection{Multiple-Particle Simulations}
In this study, we focus on single-particle images. They are simpler to simulate and provide an easily replicable geometry for measuring SNR in an experiment. However, in realistic XPIV experiment geometries, particles will be seeded at much higher densities, which would increase the image noise intensity due to scattering from the particles to the detector. We examine this effect by simulating multiple particles on a glass slide. Here, we did not investigate the effects of overlap because the likelihood of photon scattering from one particle to another, then having enough energy to be detectable, is extremely low. To confirm this, we ran a simple MCNP simulation with a 30keV pencil beam pointed through the center of two tracer particles in a vacuum. The two particles are separated by 1~$\mu$m. Photon scatter between particles and to the detector should be much more prevalent under these circumstances than normal XPIV seeding conditions. We choose a 30keV beam because that is the average photon energy of our polychromatic 55kV source spectrum. After simulating 1.2 million photons, we found the photon counts from MCNP to be only 4.4\% lower than BL (we neglected multi-detection for these simulations). For overlapping particles, direct transmission attenuation is the primary phenomenon for simulating overlapped particle images. Both BL and MCNP simulate this phenomenon accurately. Furthermore, we did not consider overlapping particles in order to isolate the effect that multiple particles' scatter to the detector will have on image noise intensity.

\begin{figure}
    \centering
    \includegraphics[width=0.5\textwidth]{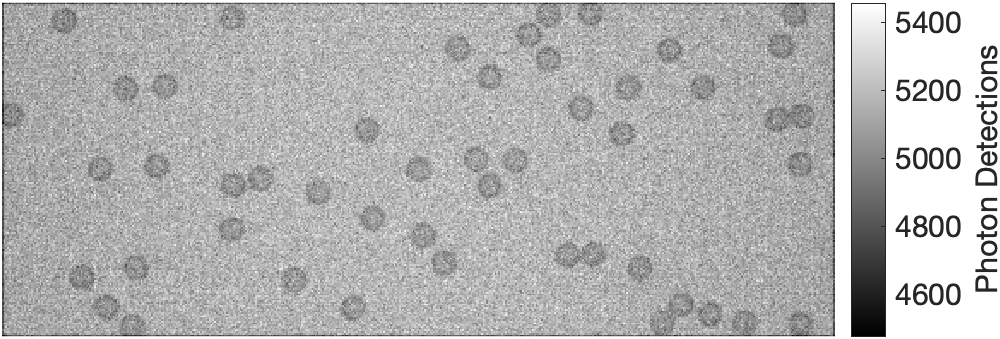}
    \caption{An MCNP-simulated image of multiple particles on a glass slide. The imaging settings are identical to the single-particle image.}
    \label{fig:mcnpmulti}
\end{figure}

\begin{figure}
    \centering
    \includegraphics[width=0.5\textwidth]{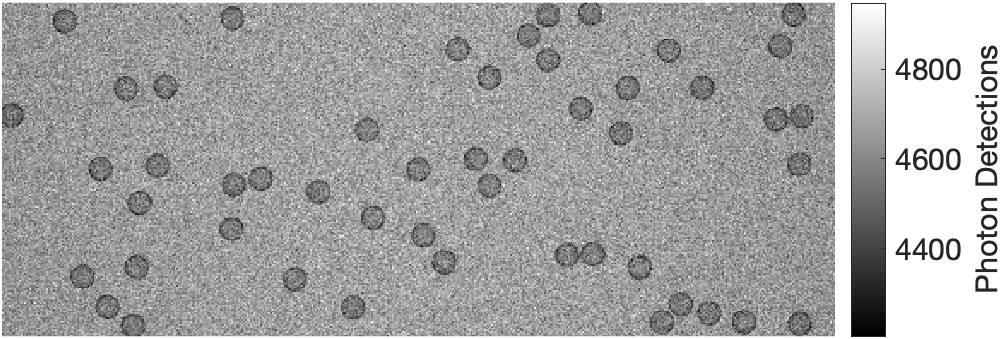}
    \caption{The BL-simulated image of multiple particles on a glass slide. As for the MCNP simulation, the settings are identical to the single-particle simulations.}
    \label{fig:blmulti}
\end{figure}

To demonstrate that the effect of scatter from the particles to the detector is small, and that our simulation tools can simulate multiple-particle images, we provide MCNP and BL images of multiple particles on a glass slide, seen in figures \ref{fig:mcnpmulti} and \ref{fig:blmulti}. The particle images cover approximately 10\% of the detector imaging area, a typical percentage for low density seeding applications that may be more likely to be used in early XPIV experiments.

To calculate the SNR, we use a smaller background area compared to the single-particle simulations in order to avoid including pixels with other particle images. Using identical background areas and particles, the background noise intensity in the multi-particle MCNP simulation increases approximately 4\% compared to the single-particle MCNP simulations; the contrast decreases by roughly 1\%. The SNR decreases by 5\%. In the multi-particle BL simulation, the SNR changes by less than 1\%. The background noise intensity is constant at 68 detections. The contrast increases from 106 to 107 detections from the single-particle simulation to the multi-particle simulation. As expected, the BL simulations are insensitive to multiple particles in the domain because it cannot simulate scatter.

In realistic imaging scenarios, scatter is dominated by the surrounding liquid, containers, and mounting equipment. Hence, the observed impact of multiple particles on noise intensity and contrast are likely exaggerated in the multi-particle simulations, where the tracer particles comprise a larger percentage of the geometry than they likely would in a early in-lab XPIV experiments. These results imply that a single-particle simulation is an adequate simplification for estimating SNR for practical design of experiments purposes. In particular, single particle simulations are adequate for comparing tracer particle designs.

\section{Conclusions}
Two approaches to predict X-ray images were explored: BL ray tracing and MCNP photon tracking. Of the two, MCNP is marginally more accurate because it models the physics better. However, MCNP is also significantly more computationally expensive than even the non-optimized BL code. BL is able to predict system behavior effectively to the same degree as MCNP at a fraction of the computational cost. Both simulation techniques predict the detectability of the AGSF-33 tracer particles with an SNR greater than unity with laboratory equipment; these predictions are confirmed by experiment images. Combined, these results imply that AGSF-33 particles could be used for laboratory-scale XPIV.

The capacity of BL simulations to cheaply predict SNR behavior for X-ray images makes it easier for XPIV system designers to select X-ray sources, exposure times, tracer particles, and source settings intelligently. XPIV system designers could also compare different tracer particles to evaluate if new coatings or particle materials will improve the image SNR.

The 55~kV, 500~$\mu$A, 100~ms MCNP simulation output all of the photon track data in 6.9~hrs. Then, we completed processing the MCNP output data in MATLAB in 10.5~hrs for a total MCNP computation time of 17.4~hrs. The 55~kV, 500~$\mu$A, 100~ms non-optimized, MATLAB-based BL simulation ran in 0.56~hrs on the same workstation that the MCNP data was processed on. In spite of the higher computational cost, MCNP may still be helpful. For geometries where phenomena such as focal spot blurring, scattering, and X-ray fluorescence are more dominant effects, using MCNP simulations may still be prudent despite the higher computational cost.

MCNP simulations may be difficult to improve upon besides using better source spectrum and flux approximations specific to the experiment. MCNP is also agnostic to time, so modelling photon counting dead time and pileup should be done after the MCNP simulation by assigning event times based on the statistics. Without such an addition, MCNP is incapable of predicting non-Poisson image noise. BL simulations, on the other hand, have ample room for improvement with regard to noise modeling, source spectrum and flux modeling, and quantitative blurring. Furthermore, the BL code for this study is far from optimized. An optimized and parallelized GPU code will only add to the computational expense advantage that BL has over the already-optimized MCNP code. Efforts to improve BL may be more fruitful than trying to use MCNP as an XPIV design tool. In all but the most extreme cases it is hard to justify the need to use MCNP for the design of XPIV experiments.

The data presented in this study shows that X-ray images of AGSF-33 particles on a glass slide show the particles clearly and with image quality adequately predicted by the models. Using SNR as a performance metric, we have the ability to methodically simulate XPIV images to improve XPIV systems with less of a trial-and-error approach. In a follow up study we will demonstrate quantitative in-lab XPIV of a flow. XPIV has great promise for rapid improvement in coming years thanks to rapidly improving system components.  As new PCDs are released with higher count rates, frame rates, and more energy thresholds, new measurements are possible. With finer energy resolution, for example, one can determine the materials in the flow, making particle detection even easier and enabling new noise and artifact suppression techniques. Additionally, techniques that are common in standard PIV, such as computed tomography (CT) reconstruction, can be easily extended to XPIV. The first generation of XPIV systems may be limited to two-dimensional projections, but CT reconstruction will enable full three-dimensional velocimetry. Besides advancements in PCD technology, laboratory X-ray sources are also steadily improving. X-ray sources with liquid metal targets \cite{stock_liquid-metal-jet_2014} are able to generate much higher photon flux and are becoming more common in laboratory settings.

The techniques developed herein can be used to harness these improvements. As laboratories adopt XPIV as a measurement tool, they can apply BL or MCNP simulation methods to quantify the expected performance of their source, tracer particle, and imager systems for a variety of flows.

\section*{Acknowledgements}
We gratefully acknowledge the support of NSF EAGER award \#1922877 program manager Ron Joslin and Shahab Shojaei-Zadeh. The contributions of Angel Rodriguez who helped with the X-ray facility setup are also gratefully recognized. 

\bibliographystyle{unsrt}
\bibliography{arxiv_version.bib}

\begin{appendices}
\section{BL Simulation Validation}\label{app:A}
\subsection*{Depth and Attenuation Validation}
The BL equation is evaluated along rays that extend from the source to the center of a sub-pixel. It is implicitly assumed that the photon flux across a sub-pixel is homogeneous.

The first validation test was conducted by calculating the true depth and attenuation ratio by hand. A hand calculation for an individual pixel was compared to the simulation results. Then, the monochromatic photon attenuation ratio and the polychromatic intensity spectrum were calculated by hand to compare to simulation. Both the depth and attenuation errors can be attributed to computational round-off.

An 86~mm $\times$ 86~mm pane of glass with a depth of 172~$\mu$m is placed 1~cm from the point source such that the upper left corner aligns with the origin. The ray from the source to the pixel (270,122) passes through the front and back faces of the glass pane. The path length of the ray passing through the glass can be calculated by
\begin{equation}\label{eq:pathlength}
    l = \frac{t}{\cos \theta \cos \phi}
\end{equation}
where $t$ is the glass pane thickness, $\theta$ is the angle between the ray and the axis pointed in the thickness direction, and $\phi$ is the angle between the ray and the direction perpendicular to the thickness direction. The difference between the hand-calculated depth and the simulation-calculated depth for the pixel (270,122) is $-3.429 \times 10^{-12}$~m.

\begin{figure}
    \centering
    \includegraphics[width = 0.5\textwidth]{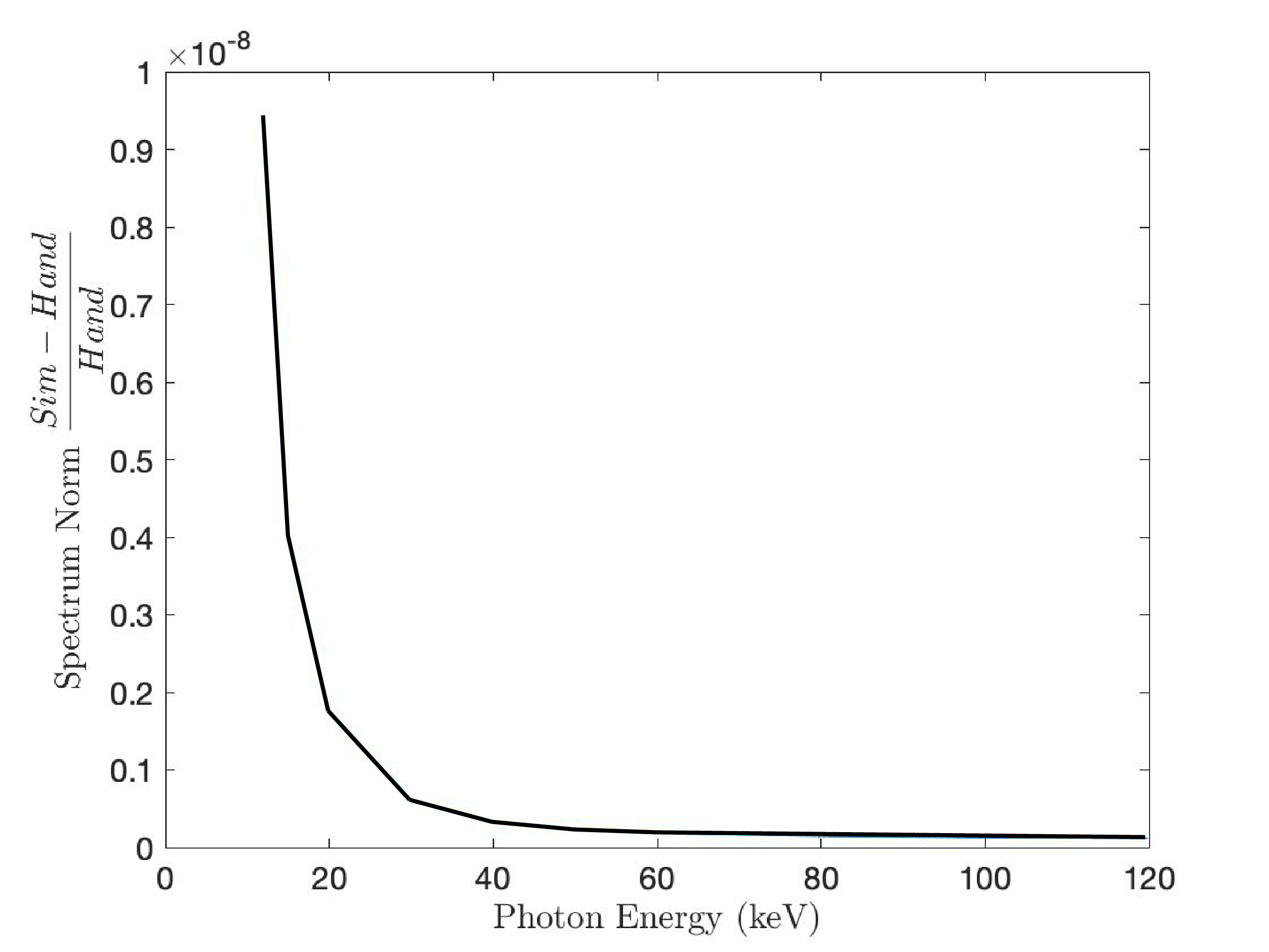}
    \caption{The error between the hand-calculated polychromatic spectrum and the MATLAB-simulated spectrum using our BL code is less than one millionth of a percent. (At low photon energies, the number of photons at that energy approaches zero, hence the singularity.)}
    \label{fig:polychromerror}
\end{figure}

For the same (270,122) pixel, the error between the hand-calculated (using path length calculated above) and simulated attenuation ratio $\frac{I}{I_0}$ is $-6.711 \times 10^{-11}$. From figure \ref{fig:polychromerror} it is clear that the error between the polychromatic spectrum calculated by hand and with the simulation is order of round off error.

\subsection*{Sub-pixel Homogeneity}
\begin{figure}
    \centering
    \includegraphics[width=0.5\textwidth]{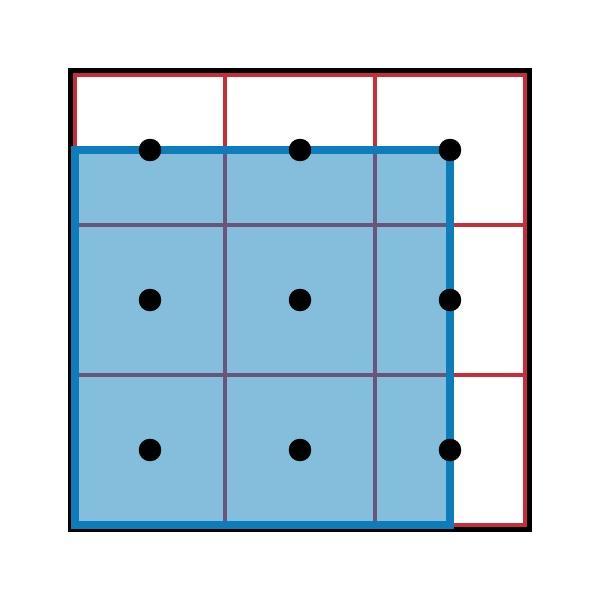}
    \caption{The scenario with the maximum lost area given a continuous, non-convex shape with dimensions greater than or equal to a pixel side-length. A sub-pixel is assumed to be uniformly covered or uncovered based on the state of its center.}
    \label{fig:BLhom1}
\end{figure}

In order to validate the homogeneity assumption (i.e. that a sufficient number of sub-pixels was considered), an upper bound on the maximum ``lost area" is established. Lost area is the area of a pixel that is erroneously attributed as covered (or not covered). For example, in Figure \ref{fig:BLhom1}, the lost area is the white region. The sub-pixels detect the blue material projection, but the projection does not physically cover the entire pixel. Assuming a continuous, non-convex, projected shape whose dimensions are larger than a pixel side length, the maximum lost area can be readily calculated. The scenario depicted in Figure \ref{fig:BLhom1} is the maximum possible lost area that meets the stated assumptions. The area lost as a fraction of the pixel area is given as a function of the number of sub-pixels, $N$, by
\begin{equation}\label{eq:lostA1}
    \frac{ A_{lost} }{ A_{pixel} } = \frac{1}{\sqrt{N}} - \frac{1}{4N}.
\end{equation}
A pixel is sufficiently homogeneous when the maximum lost area is no more than 10\% of the pixel area. For $N=121$ sub pixels (11$\times$11), the maximum lost area is less than 9$\%$.

\begin{figure}
    \centering
    \includegraphics[width=0.5\textwidth]{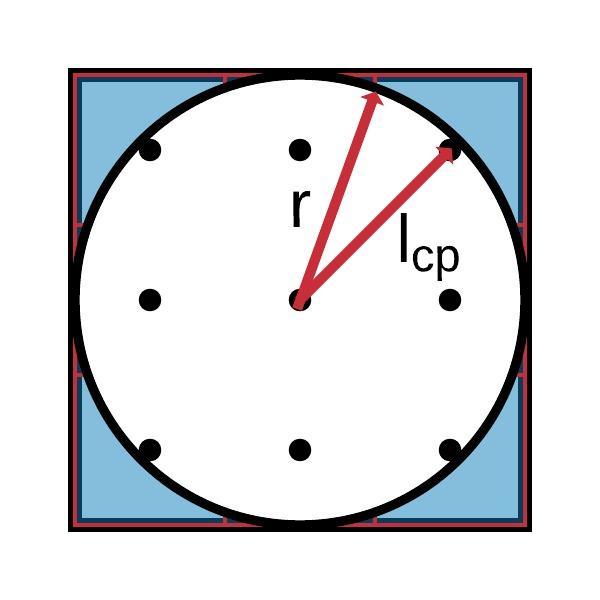}
    \caption{The maximum lost area scenario for a hole with a diameter greater than or equal to the pixel side length.}
    \label{fig:BLhom2}
\end{figure}

\begin{figure}
    \centering
    \includegraphics[width = 0.5\textwidth]{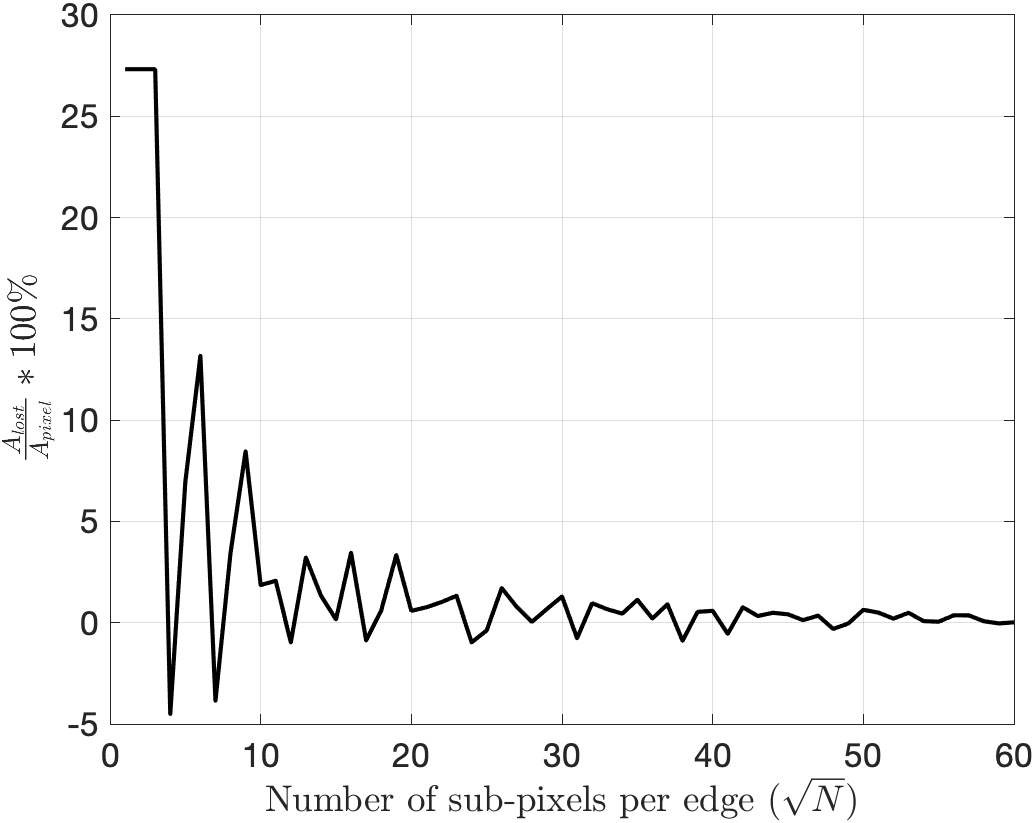}
    \caption{As the number of sub-pixels increases, the maximum area lost decreases rapidly. The homogeneity assumption criteria is met for $N \geq 121$ (11$\times$11).}
    \label{fig:circAreaConv}
\end{figure}

If the non-convex assumption is relaxed to allow for holes, the maximum lost area bound increases. Figure \ref{fig:BLhom2} shows the maximum possible lost area after relaxing the non-convex assumption. The lost area was calculated in MATLAB; the results are plotted in figure \ref{fig:circAreaConv}. It is clear that the criteria for the homogeneity assumption, discussed above, is again met for $N \geq 121$ (11$\times$11).

\section{MCNP PTRAC File Explanation}\label{app:B}
MCNP \cite{cj_werner_et_al_mcnp62_2018} with PTRAC files is a powerful simulation tool, but care must be taken when analyzing the data as the PTRAC file format is not intuitive. We will briefly make a few comments on this so that anyone interested on building on this this study may better navigate some of the challenges.

The PTRAC file denotes different types of interactions – henceforth referred to as events – and the associated event data: location, direction, and post-event photon energy. The physically relevant event types are: surface, collision, and termination events (sur, col, and ter, respectively). Each event has an associated event ID, which is a number reported in the PTRAC file. A surface event flags a photon entering the CdTe detector panel; its event ID is 3000. With the filters discussed prior, a surface event flags a photon entering the detector panel. A collision event flags a photon imparting some energy to the detector; its event ID is 4000. Collision events can refer to either a scattering event or an absorption event. Termination events flag the end of a photon's recorded track history; its event ID is 5000. When a photon is absorbed, identical collision and termination events are reported in the PTRAC file. In addition to the physical event types, there is a fourth event type: history termination events. Their event IDs are 9000. History termination events denote the end of recording a photon track. This may occur when a photon is absorbed or when it leaves the domain of interest (in this case, the detector panel). When a photon is absorbed, a history termination event identical to the collision and termination events is recorded.

\begin{figure}[h!]
    \onecolumn
    \centering
    \includegraphics[width=\textwidth]{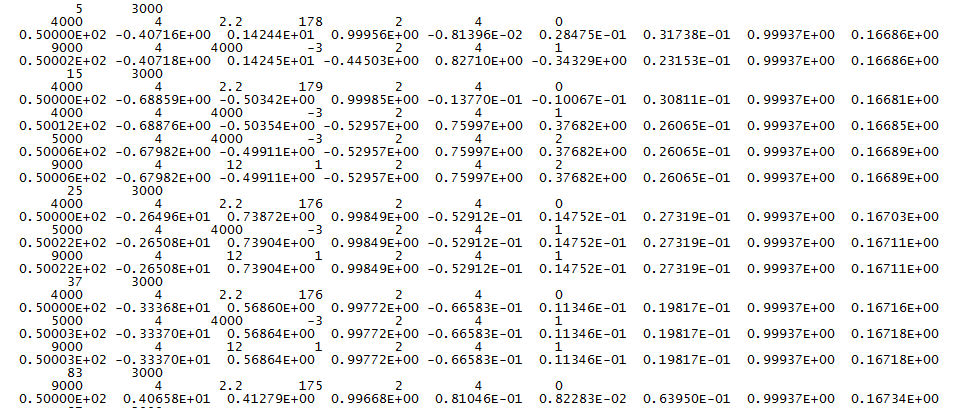}
    \caption{Example particle history tracks from an MCNP PTRAC file.}
    \label{fig:ptrac}
\end{figure}

Figure \ref{fig:ptrac} shows a snippet of the PTRAC file from one of our trial MCNP simulations. In this example, the first particle track is for the fifth source particle. As with every particle track in this PTRAC file, the history begins with a surface event. The ``5   3000" line at the very beginning states that the initial event for the track history of source particle five is a surface entry event. The data associated with this event can be seen on the third line. The first three numbers on line three are the x-, y-, and z-coordinates in centimeters; the next three numbers are the direction of travel component vectors after the event occurs. The last three numbers are photon energy in MeV, particle weighting, and track time in shakes.

In the case of this first particle track, the photon enters the detector at (50, -0.40716, 1.4224) centimeters with 0.31738~MeV of energy. The photon then collides (event ID 4000 on line two) at (50.002, -0.40718, 1.4245)~cm and departs with 0.23153~MeV of energy. Note that the particle direction of travel is reversed in the x-direction due to this collision; the particle leaves the detector (event ID 9000 on line 4) before being absorbed, i.e., terminated (event ID 5000). In this track history, the photon deposits 0.08585MeV of energy in the detector.

\pagebreak
\onecolumn

\end{appendices}

\end{document}